\documentclass[aps,showpacs,floatfix,twocolumn,superscriptaddress,longbibliography]{revtex4-2}

\usepackage{amsmath}
\usepackage{amssymb}
\usepackage{amstext}
\usepackage{amsopn}
\usepackage{amsfonts}
\usepackage{amsxtra}
\usepackage[english]{babel}
\usepackage{graphicx}
\usepackage{float}
\usepackage{bm}
\usepackage{multirow}
\usepackage{mathrsfs}
\usepackage{mathtools}
\usepackage{dcolumn}
\usepackage[colorlinks]{hyperref}
\usepackage{color}
\usepackage{url}
\usepackage{breakurl}
\usepackage{relsize}

\begin{document}
\title{Anisotropic optical properties of detwinned BaFe$_\mathbf{2}$As$_\mathbf{2}$}
\preprint{Draft, not for distribution}
\author{Christopher C.~Homes}
\email[]{homes@bnl.gov}
\affiliation{Condensed Matter Physics and Materials Science Division, Brookhaven National Laboratory,
  Upton, New York 11973, USA}
\author{Thomas Wolf}
\author{Christoph Meingast}
\email[]{christoph.meingast@kit.edu}
\affiliation{Institute for Quantum Materials and Technologies, Karlsruhe Institute of Technology,
76021 Karlsruhe, Germany}
%
%
\date{\today}

%
%
\begin{abstract}
The optical properties of a large, detwinned single crystal of BaFe$_2$As$_2$ have been
examined over a wide frequency range above and below the structural and magnetic transition
at $T_{\rm N}\simeq 138$~K.
Above $T_{\rm N}$ the real part of the optical conductivity and the two infrared-active lattice
modes are almost completely isotropic; the lattice modes show a weak polarization dependence
just above $T_{\rm N}$.
For $T<T_{\rm N}$, the optical conductivity due to the free-carrier response is anisotropic, being
larger along the \emph{a} axis than the \emph{b} axis below $\simeq 30$~meV; above this energy the
optical conductivity is dominated by the interband contributions, which appear to be isotropic.
The splitting of the low-energy infrared-active mode below $T_{\rm N}$ is clearly observed, and
the polarization modulation of the new modes may be used to estimate that the crystal is
$\simeq 70$\% detwinned.  The high-frequency mode, with a threefold increase in strength of
the lower branch below $T_{\rm N}$ and nearly silent upper branch, remains enigmatic.
\end{abstract}
%
%
%
\maketitle

%
%
%
\section{Introduction}
In the pantheon of iron-based superconductors, the \emph{Ae}Fe$_2$As$_2$
(``122'') materials, where \emph{Ae}$\,=\,$Ca, Sr, or Ba, are of particular
importance because of the many ways in which superconductivity may be
induced \cite{Rotter2008a,Sefat2008,Ni2008,Sasmal2008,Chen2008,Alireza2008,
Chu2009,Goko2009,Saha2009,Jiang2009,Shi2010,Cortes2011,Ishikawa2009,Colombier2009,
Kitagawa2009}.  At room temperature they are paramagnetic metals, but as the
temperature is reduced they undergo a structural transition from a tetragonal
($I4/mmm$) to an orthorhombic ($Fmmm$) unit cell, which is accompanied by a
magnetic transition and the formation of spin-density-wave-like (SDW) order where
the  moments are aligned in the \emph{a-b} planes; ferromagnetically (FM) along
the \emph{b} axis, and antiferromagnetically (AFM) along the \emph{a} axis.
In the case of BaFe$_2$As$_2$, this transition occurs at $T_{\rm N}\simeq 138$~K
\cite{Rotter2008b}.
In the orthorhombic phase, the crystals are heavily twinned.  Early optical
studies of the in-plane optical properties of BaFe$_2$As$_2$ examined the average
of both orientations in the SDW state \cite{Hu2008,Akrap2009,Pfuner2009}; however,
the application of uniaxial stress along the (110) direction for the tetragonal
unit cell results in a nearly twin-free sample \cite{Tanatar2009,Fisher2011}.
Transport \cite{Chu2010,Kissikov2018}, as well as optical measurements \cite{Nakajima2011,Dusza2012,Mirri2015},
of detwinned samples below $T_{\rm N}$ reveal an anisotropic response where the
resistivity along the AFM direction is lower than that along the FM direction.
This is a counterintuitive result as one would normally expect that that the
scattering from spin fluctuations would result in a higher resistivity along the AFM
direction \cite{Chu2010}.  On the other hand, this anisotropy is sharply reduced in
annealed samples for $T\ll T_{\rm N}$ \cite{Nakajima2011}, and the resistivity is
indeed observed to be higher along the AFM direction in the related material FeTe
below $T_{\rm N}$, a result attributed to Hund's rule coupling making transport along
the FM direction easier than along the AFM direction \cite{Jiang2013}.
While the designs of the clamped cells used to mechanically detwin single crystals for
optical studies are quite elegant \cite{Nakajima2011,Dusza2012}, this approach
necessarily requires that the size of the imaging spot is smaller than the crystal,
leading to a reduced signal and limiting the ability to track weak spectral
features such as lattice modes.

%
%
\begin{figure}[tb]
\centerline{\includegraphics[width=3.1in]{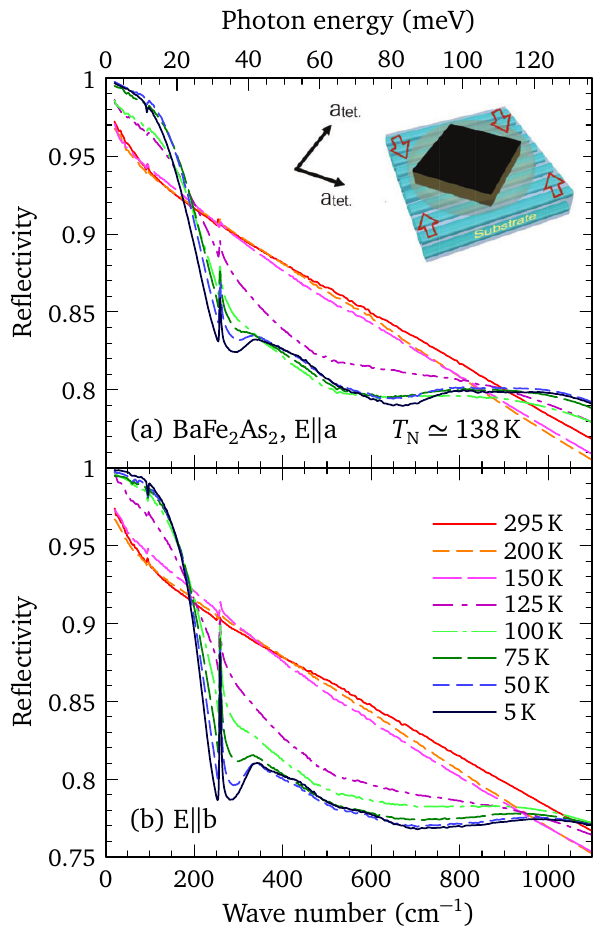}}%
\caption{The reflectivity in the infrared region for a
single crystal of BaFe$_2$As$_2$ at several temperatures above and below
$T_{\rm N}\simeq 138$~K for: (a) light polarized along the \emph{a} axis;
(b) light polarized along the \emph{b} axis.  The resolution at low frequency
is typically better than 2~cm$^{-1}$.  The inset shows the orientation of the
GFRP with respect to the tetragonal unit cell.}
\label{fig:reflec}
\end{figure}

%
%
In this work we examine the temperature dependence of the optical properties of
a large single crystal of BaFe$_2$As$_2$ that has been detwinned through the application
of a symmetry-breaking strain, based on differential thermal expansion \cite{He2017}.
While previous studies of detwinned BaFe$_2$As$_2$ have been performed \cite{Nakajima2011,Dusza2012,Mirri2015},
the current study employs overfilling-technique \cite{Homes1993} whereby the
entire crystal face may be examined, allowing the optical conductivity and the
infrared-active modes to be studied.  The optical conductivity is essentially isotropic
for $T > T_{\rm N}$.  The relatively large size of the sample allows the two normally
infrared-active modes above $T_{\rm N}$ to be identified; the lattice modes show a
slight polarization dependence just above $T_{\rm N}$ in response to the applied strain.
For $T<T_{\rm N}$ the optical conductivity due to the free-carrier response is anisotropic
and is higher along the \emph{a} direction, in agreement with previous work.  Interestingly,
the interband contributions appear to be isotropic.  Below $T_{\rm N}$ the degeneracy
of the infrared-active vibrations is lifted and they split into two new modes that are
optically-active along either the \emph{a} or \emph{b} axis; by examining the
polarization modulation of the low-frequency modes below $T_{\rm N}$, it is possible to
 estimate that the sample is about 70\% detwinned.  The behavior of the high-frequency
mode is very curious, with one branch increasing dramatically in strength, while the
other remains largely silent.
This technique for detwinning crystals may in principle be used to allow further detailed
optical studies of, \emph{e.g.}, the Ba(Fe$_{1-x}$Co$_x$)$_2$As$_2$ family of materials.

%
%
\section{Experiment}
Large single crystals of BaFe$_2$As$_2$ were grown by a self-flux method \cite{He2017}.
The crystals have well-defined growth faces, allowing the \emph{a} axis in the
tetragonal phase to be identified.  A piece of thin glass-fiber reinforced plastic (GFRP)
$\simeq 0.5$~mm thick was cut and shaped to match a crystal approximately $2\,{\rm mm}\times 2\,{\rm mm}$
and $\simeq 100\,\mu{\rm m}$ thick, and attached using epoxy with the fibers oriented along
the (110) direction in the tetragonal phase, shown in the inset of Fig.~\ref{fig:reflec}(a).
It has been demonstrated that the difference of the thermal expansion parallel and
perpendicular to the fiber direction of the GFRP substrate is comparable to the
orthorhombic distortion in BaFe$_2$As$_2$ near $T_{\rm N}$, resulting in a large symmetry
breaking strain; this technique has been used successfully to measure the resistivity and
susceptibility anisotropies in BaFe$_2$As$_2$ (results and experimental details regarding
this method of detwinning may be found in Ref.~\onlinecite{He2017}).  The entire arrangement
was glued to the tip of an optically-black cone.  The temperature dependence of the
reflectivity has been measured above and below $T_{\rm N}$ over a wide frequency range
(2~meV to over 3~eV) using an overfilling technique in combination with {\em in situ}
evaporation \cite{Homes1993}; the results are shown in the infrared region in
Figs.~\ref{fig:reflec}(a) and \ref{fig:reflec}(b) for light polarized along the
\emph{a} and \emph{b} axes in the orthorhombic phase, respectively.
Above $T_{\rm N}$ the reflectivity along the two polarizations is nearly identical;
only the reflectivity at 150~K along the \emph{b} axis appears to be slightly higher
than its counterpart along the \emph{a} axis.
Below $T_{\rm N}$ the optical properties are strongly anisotropic.  A plasma-like
edge develops in the reflectivity for both polarizations.  For $T\ll T_{\rm N}$
the low frequency reflectivity approaches unity, while above $\simeq 20$~meV the
reflectivity decreases rapidly before forming a plateau above $\simeq 50$~meV;
however, the reflectivity levels and the width of the plasma-like edge are very
different along the \emph{a} and \emph{b} directions, in agreement with previous
optical studies of this material \cite{Nakajima2011,Dusza2012,Mirri2015}.  Additional
structure is observed in the mid-infrared region up to about 0.5~eV, above which the
reflectivity is comparable to the room-temperature values and displays little
temperature dependence (the temperature dependence of the reflectivity is shown
over a wide range in Fig.~S1 of the Supplementary Material \cite{Suplmt}).
Superimposed upon the reflectance are two sharp features attributed to the infrared-active
lattice modes at $\simeq 95$ and 256~cm$^{-1}$ \cite{Homes2018}, which display an
anisotropic response below $T_{\rm N}$.

%
%
While the reflectance is a useful quantity, it is a combination of the real and
imaginary parts of the dielectric function, and as such, is not an intuitive quantity.
The complex dielectric function, $\tilde\epsilon(\omega)=\epsilon_1+i\epsilon_2$,
has been determined from a Kramers-Kronig analysis of the reflectivity.  At low
frequency, a metallic Hagen-Rubens extrapolation, $R(\omega) = 1 - A\sqrt{\omega}$
was employed, where $A$ is chosen to match the value of the reflectance at the lowest
measured frequency.  Above the highest-measured frequency point the reflectance was assumed
to be constant to $8\times 10^4$~cm$^{-1}$, above which a free-electron approximation
($R\propto \omega^{-4}$) was assumed \cite{wooten}.

%
%
\begin{figure}[b]
\centerline{\includegraphics[width=3.1in]{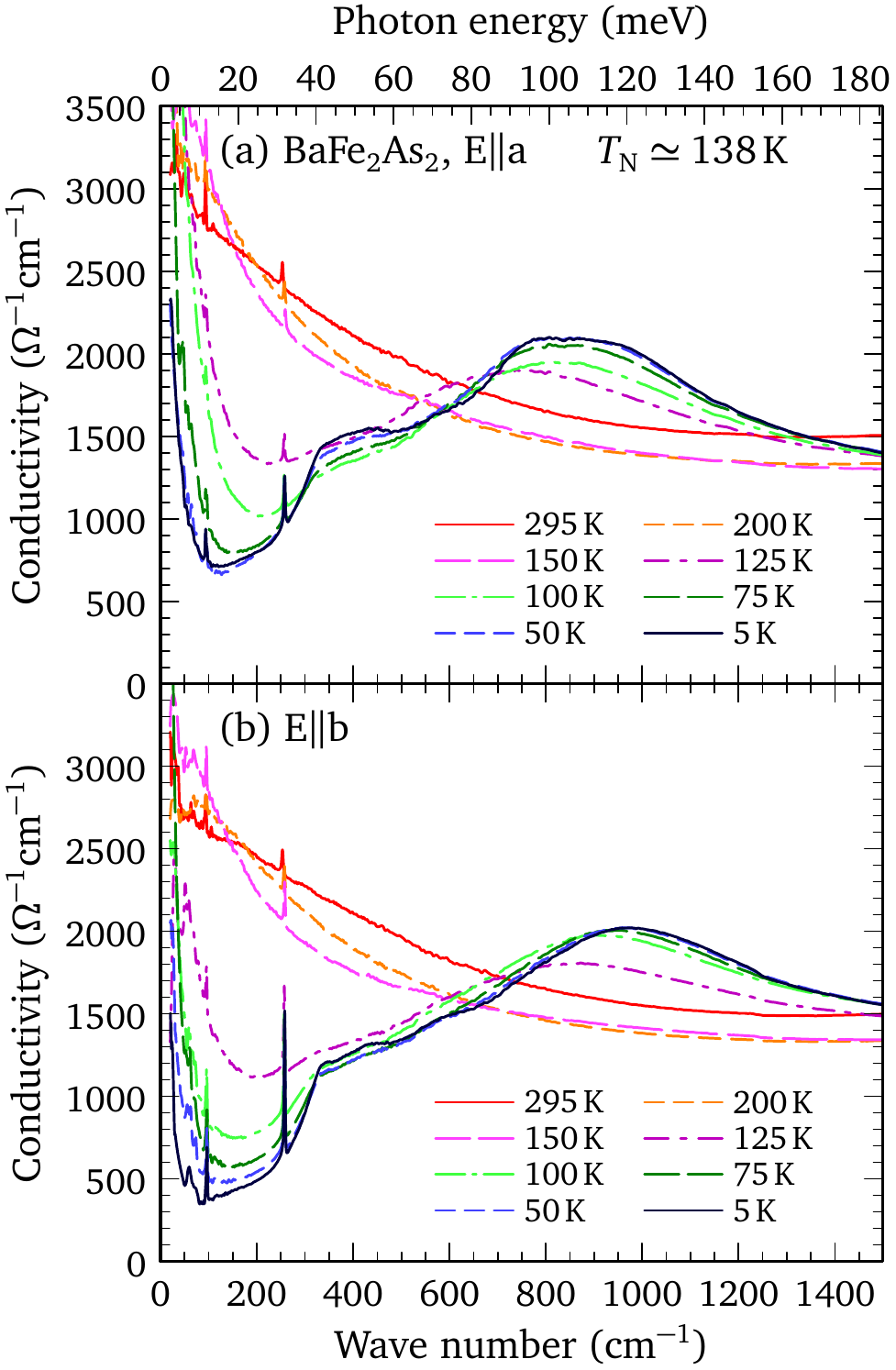}}%
\caption{The temperature dependence of the real part of the optical conductivity of
of BaFe$_2$As$_2$ above and below $T_{\rm N}$ for light polarized along (a) the
\emph{a} axis, and (b) the \emph{b} axis.}
\label{fig:sigma}
\end{figure}

\section{Results and Discussion}
The temperature-dependence of the real part of the optical conductivity is
shown for light polarized along the \emph{a} and \emph{b} axis in Figs.~\ref{fig:sigma}(a)
and \ref{fig:sigma}(b), respectively.  Above $T_{\rm N}$ the real part of the
optical conductivity for the two polarizations are almost identical; however,
below $T_{\rm N}$, there is a remarkable anisotropy below $\simeq 300$~cm$^{-1}$
where the conductivity along the \emph{a} direction is larger than along \emph{b},
in agreement with other studies \cite{Nakajima2011,Dusza2012,Mirri2015}.
In addition to the broad features associated with the free-carrier response and the
interband excitations, there are narrow lattice modes which also display an anisotropic
response.  The behavior of the electronic properties will be examined first, followed by
the lattice modes.

%
%
\subsection{Electronic response}
The optical properties of this multiband material have been studied extensively in the
twinned materials and are described by a Drude-Lorentz model where at least two different
contributions to the free-carrier response are considered \cite{Wu2010}, resulting in the
complex dielectric function,
%
%
\begin{equation}
  \tilde\epsilon(\omega) = \epsilon_\infty - \sum_j{{\omega_{p,D;j}^2}\over{\omega^2+i\omega/\tau_{D,j}}}
    + \sum_k {{\Omega_k^2}\over{\omega_k^2 - \omega^2 - i\omega\gamma_k}},
  \label{eq:dl}
\end{equation}
where $\epsilon_\infty$ is the real part of the dielectric function at high
frequency, $\omega_{p,D;j}^2 = 4\pi ne^2/m^\ast$ and $1/\tau_{D,j}$ are the
plasma frequency and scattering rate for the delocalized (Drude) carriers for
the $j$th band, respectively; $\omega_k$, $\gamma_k$ and $\Omega_k$ are the
position, width, and oscillator strength of the $k$th vibration or bound excitation
(the intensity is proportional to $\Omega_k^2$).
The complex conductivity, $\tilde\sigma(\omega)$, is calculated from from the complex
dielectric function, $\tilde\sigma(\omega) = \sigma_1 +i\sigma_2 = -2\pi i \omega
[\tilde\epsilon(\omega) - \epsilon_\infty ]/Z_0$, where $\epsilon_\infty$ is the high-frequency
contribution to the real part of the dielectric function, and $Z_0\simeq 377$~$\Omega$ is
the impedance of free space.  The real and imaginary parts of the complex conductivity
have been fit simultaneously to Eq.~(\ref{eq:dl}) using a non-linear least-squares technique.

%
%
\begin{table}[tb]
\caption{The plasma frequencies and scattering rates of the narrow (D1) and broad
(D2) Drude terms, as well as several Lorentzian components, returned from the fits
to the real and imaginary parts of the optical conductivity for BaFe$_2$As$_2$ for
light polarized in the \emph{a-b} planes at $T\gtrsim T_{\rm N}$, and for light
polarized along the \emph{a} and \emph{b} axes at $T\ll T_{\rm N}$. All units are in
cm$^{-1}$.}
\begin{ruledtabular}
\begin{tabular}{c l c c c c}
  Pol.   &  Temperature & Component &   $\omega_k$ & $1/\tau_D$, $\gamma_k$ & $\omega_{p,D}$, $\Omega_k$ \\
\cline{1-6}
%
%
 E$\parallel${ab} & ($T\simeq T_{\rm N})$ & D1 &  --  &  130 & 4110 \\
                  &                       & D2 & --   & 1346 & $11\,900$ \\
                  &                       & LM & 4052 & 6110 & $23\,004$ \\
 E$\parallel${a}  & ($T\ll T_{\rm N})$    & D1 &  --   &  3.6 & 3450 \\
                  &                       & D2 &  --   &  146 & 2910 \\
                  &                       & L1 &  361  &  293 & 3900 \\
                  &                       & L2 &  861  &  989 & 10196 \\
                  &                       & LM & 4154  & 6500 & 24036 \\
 E$\parallel${b}  & ($T\ll T_{\rm N})$    & D1 & --    &  2.1  & 3690 \\
                  &                       & D2 & --    &  207  & 2080 \\
                  &                       & L1 &  387  &  326  & 3900 \\
                  &                       & L2 &  953  & 1110  & 10382 \\
                  &                       & LM & 4012  & 6258  & 24584 \\
\end{tabular}
\end{ruledtabular}
\label{tab:fits}
\vspace*{-0.5cm}
\end{table}
%

%
%
Above $T_{\rm N}$, the reflectivity and the optical conductivity show
little polarization dependence. Although the features are rather broad, just above
$T_{\rm N}$ at 150~K, the fits to the optical conductivity yield a narrow Drude
component (D1), $\omega_{p,D1}\simeq 4100$~cm$^{-1}$ and $1/\tau_{D1}\simeq 130$~cm$^{-1}$,
and a broad Drude term (D2), $\omega_{p,D2}\simeq 11\,900$~cm$^{-1}$ and $1/\tau_{D2}\simeq
1300$~cm$^{-1}$ (Table~\ref{tab:fits}) in good agreement with the values from a previous
study \cite{Homes2016}.

\begin{figure}[t]
\centerline{\includegraphics[width=3.1in]{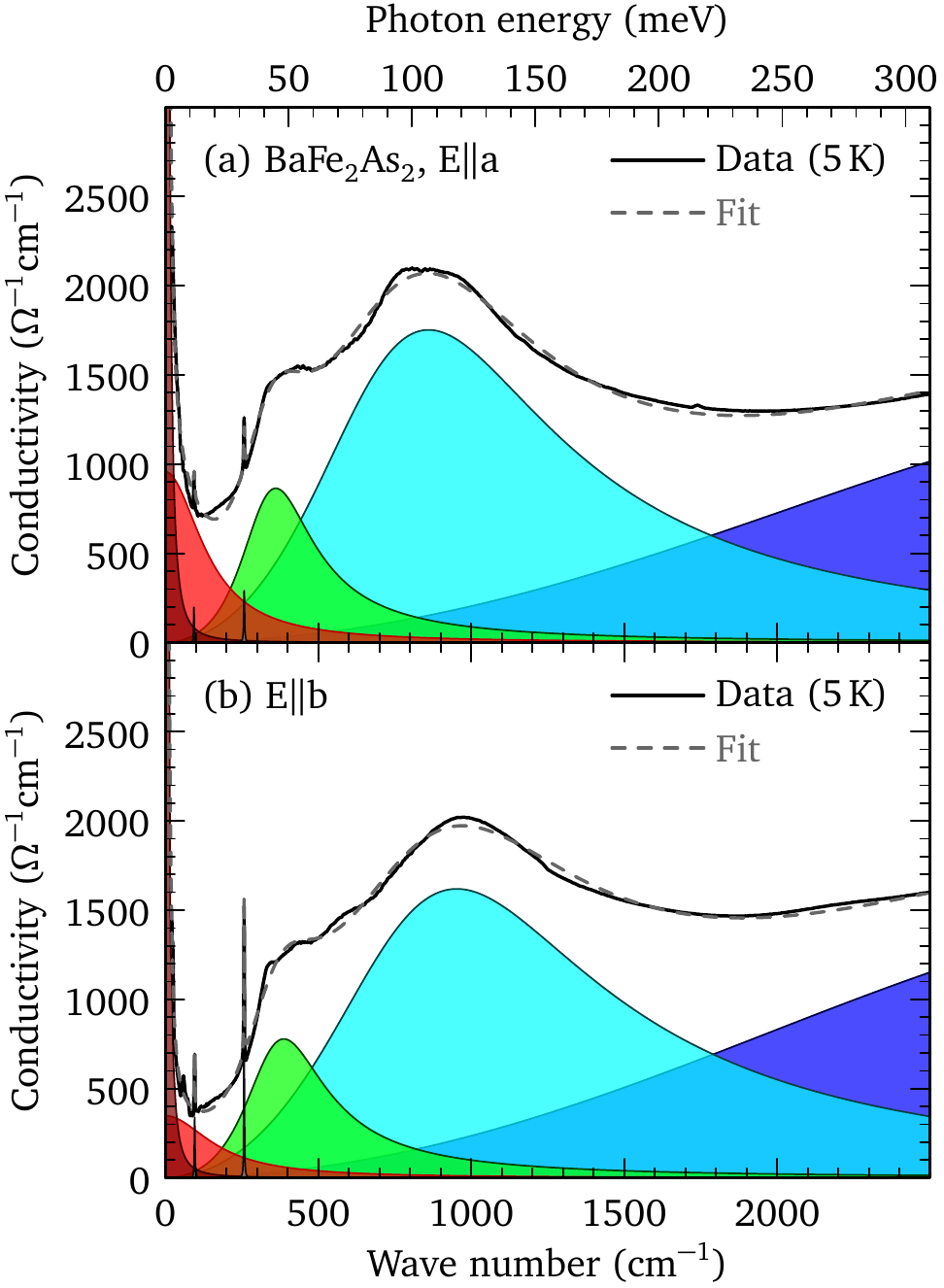}}%
\caption{The results of the fits to the complex conductivity of BaFe$_2$As$_2$ below
$T_{\rm N}$ at 5~K compared to the real part of the optical conductivity in the far- and
mid-infrared regions for light polarized along the: (a) \emph{a} axis; (b) \emph{b} axis.
The anisotropic response is strongest below $\simeq 30$~meV.  The fit is decomposed
into the contributions from the narrow and broad Drude components, as well as several
Lorentz oscillators (Table~\ref{tab:fits}). }
\label{fig:fits}
\end{figure}

%
%
Below the structural and magnetic transition at $T_{\rm N}$, the optical conductivity
undergoes significant changes, shown in Figs.~\ref{fig:reflec} and \ref{fig:sigma},
due to the reconstruction of the Fermi surface \cite{Richard2010}.  The fits for
$T\ll T_{\rm N}$ at 5~K for the \emph{a} and \emph{b} polarizations are shown in
Figs.~\ref{fig:fits}(a) and \ref{fig:fits}(b), respectively, where they have been
decomposed into the individual contributions from the various Drude and Lorentz
components.
The observed optical anisotropy of $\sigma_{1,a}/\sigma_{1,b} \simeq 2$ in the far-infrared region
is in good agreement with another study that employed a mechanical apparatus to detwin the
crystal \cite{Nakajima2011}.  As in the twinned materials, below $T_{\rm N}$ new features
appear at $\simeq 350$ and 900~cm$^{-1}$ \cite{Homes2016}; interestingly, there appears
to be little or no anisotropy in either these or other bound excitations associated
with the interband transitions in this compound.  Indeed, the oscillator parameters for
the bound excitations in Figs.~\ref{fig:fits}(a) and \ref{fig:fits}(b) are nearly
identical for both polarizations and are similar to what is observed in the twinned
material.  The anisotropy in the far-infrared region of the optical conductivity arises
purely from the behavior of the free carriers.

%
%

As in the case of the twinned materials, for $T\ll T_{\rm N}$ the plasma frequency for
the narrow Drude component undergoes only a small decrease, while the scattering rate drops
precipitously: along the \emph{a} axis, $\omega_{p,D1}\simeq 3500$~cm$^{-1}$ and
$1/\tau_{D1}\simeq 3.6$~cm$^{-1}$; along the \emph{b} axis, $\omega_{p,D1}\simeq 3700$~cm$^{-1}$
and $1/\tau_{D1}\simeq 2.1$~cm$^{-1}$.  The uncertainties associated with the small
scattering rates, and the similarity of the plasma frequencies, suggest that the narrow
Drude component is fairly isotropic below $T_{\rm N}$.
%
%
The plasma frequency for the broad Drude component decreases significantly, while the
change in the scattering rate, while significant, is not as dramatic as it is for the
narrow Drude component: along the \emph{a} axis, $\omega_{p,D2}\simeq 2900$~cm$^{-1}$ and
$1/\tau_{D2}\simeq 146$~cm$^{-1}$; along the \emph{b} axis, $\omega_{p,D2}\simeq 2100$~cm$^{-1}$
and $1/\tau_{D2}\simeq 210$~cm$^{-1}$ (Table~\ref{tab:fits}).  The large difference in
the plasma frequencies arises from a smaller effective mass along the \emph{a} axis
where the SDW is present, suggesting a decrease in the electronic correlations \cite{Yin2011a,Dai2014},
as opposed to the \emph{b} axis, or FM direction, where the larger effective mass significantly
reduces the plasma frequency, and subsequent contribution to the optical conductivity in
the far-infrared region.
%
%

%
%
\subsection{Lattice modes}
In the high-temperature tetragonal phase, the irreducible vibrational
representation for the infrared modes yields $2A_{2u}+2E_u$ vibrations
\cite{Litvinchuk2008}, where the singly-degenerate $A_{2u}$ modes are
active along the \emph{c} axis, and the doubly-degenerate $E_u$ modes are
active in the planes.  Below the tetragonal to orthorhombic structural
transition at $T_{\rm N}$, the degeneracy of the in-plane modes is
lifted, $E_u \rightarrow B_{2u} + B_{3u}$, where the $B_{2u}$ and
$B_{3u}$ modes are active along the \emph{b} and \emph{a} axes, respectively.
%
%
\begin{figure}[b]
\centerline{\includegraphics[width=3.1in]{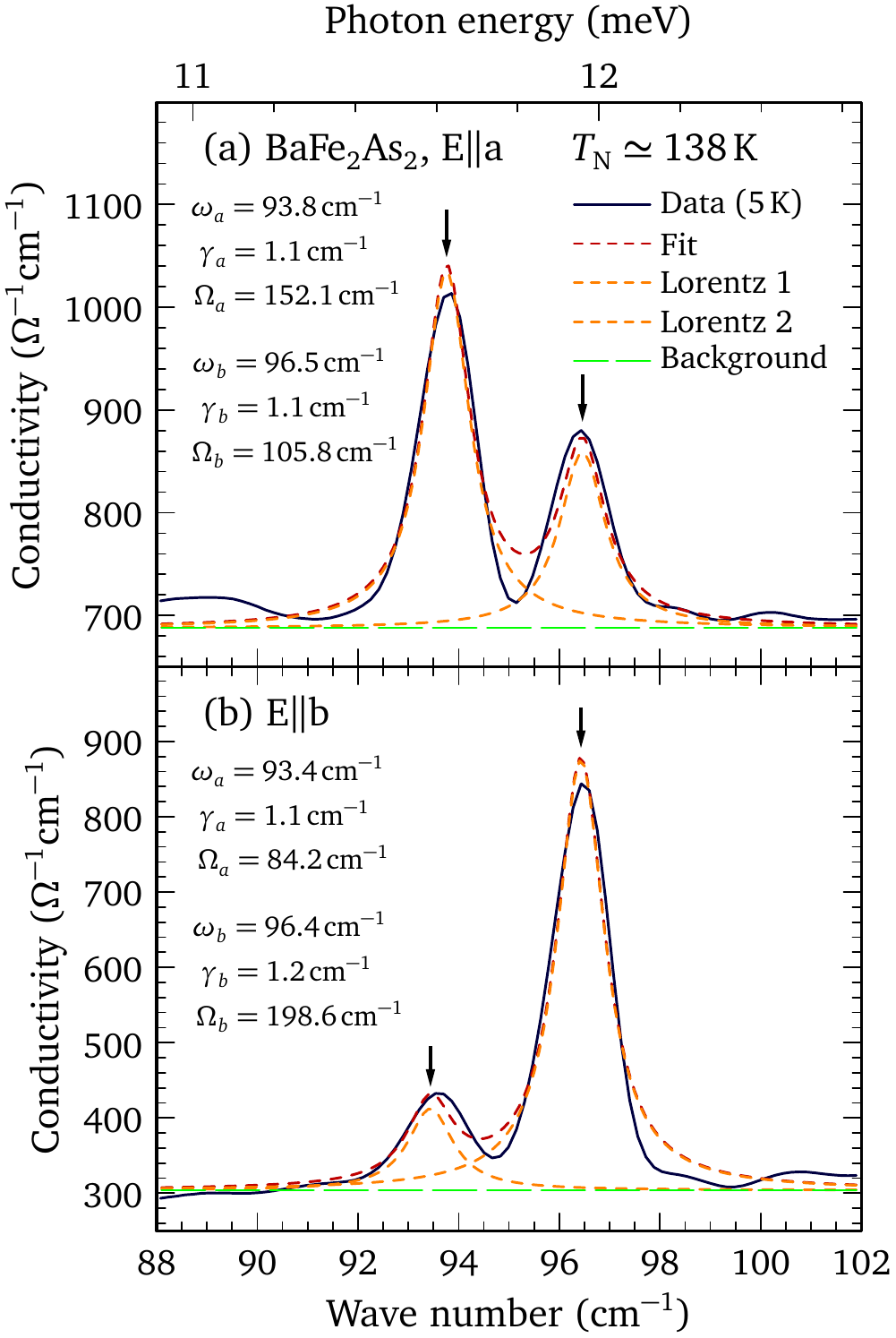}}%
\caption{The real part of the optical conductivity of BaFe$_2$As$_2$ in the orthorhombic
phase at 5~K in the region of the low-frequency $B_{3u}$ and $B_{2u}$ modes with an experimental
resolution of 1~cm$^{-1}$ for light polarized along the: (a) \emph{a} axis; (b) \emph{b} axis.
For each polarization the vibrational structure has been fit to two Lorentz oscillators
superimposed on a linear background. The fitted oscillator parameters are listed in each
panel; the positions indicated by the arrows.}
\label{fig:eu1}
\end{figure}
In the twinned samples the low-frequency $E_u$ mode observed at $\simeq 94$~cm$^{-1}$
above $T_{\rm N}$ involves the in-plane displacements of the Ba atom moving in opposition
to the Fe and As atoms; below $T_{\rm N}$ this mode splits into two features at
$\simeq 93$ and 96~cm$^{-1}$ \cite{Schafgans2011}.
The high-frequency mode at $\simeq 256$~cm$^{-1}$ does not appear to split below
$T_{\rm N}$, instead displaying an anomalous increase in oscillator strength \cite{Akrap2009};
this mode involves only the in-plane displacements of Fe and As atoms which move
in opposition to one and other \cite{Sandoghchi2013,Homes2018}.
In the detwinned samples, the low-frequency mode has not been examined; the
high-frequency mode displays the same anomalous increase in oscillator strength
below $T_{\rm N}$, but it also does not appear to split and is active only
along the \emph{b} axis \cite{Nakajima2011}.
In this work, we are able to examine the splitting and the polarization
dependence of the low-frequency $E_u$ mode below $T_{\rm N}$, and as well
as the details of the high-frequency mode.

\subsubsection{Low-frequency mode}
The low-frequency $E_u$ mode has been fit using the Lorentzian oscillator
described in Eq.~(\ref{eq:dl}) superimposed on a linear background.
Above $T_{\rm N}$ at 295~K, this mode is almost completely isotropic, with
$\omega_0\simeq 93.7$, $\gamma_0\simeq 2.5$, and $\Omega_0\simeq 195$~cm$^{-1}$.
(Asymmetric line shapes were also considered in the supplementary material, but
the asymmetry parameter was quite small, effectively resulting in a symmetric
Lorentzian oscillator.)
Interestingly, just above $T_{\rm N}$ at 150~K, there is a slight polarization
dependence with $\omega_0\simeq 94$~cm$^{-1}$ along along the \emph{a} axis, and
$\omega_0\simeq 95$~cm$^{-1}$ along the \emph{b} axis, while the width and
strength show no such dependence.  This indicates that GFRP is imparting some strain
on the crystal and creating a slight asymmetry just above $T_{\rm N}$.
This is consistent with the observation of a significant anisotropy in the
resistivity just above $T_{\rm N}$, which was attributed to the magnetic transition
rather than nematic fluctuations \cite{He2017}.  Similarly, the vibrational splitting
just above the transition would suggest that the phonons are coupling to the magnetism.
Below $T_{\rm N}$ this vibration clearly splits into two modes at $\simeq 94$ and 96~cm$^{-1}$
at low temperature, both of which display a strong polarization dependence.  The
optical conductivity is shown in the region of the low-frequency $B_{3u}$ and $B_{2u}$
modes at 5~K for light polarized along the \emph{a} and \emph{b} axis, in
Figs.~\ref{fig:eu1}(a) and \ref{fig:eu1}(b), respectively; the different
contributions are denoted in the legend, while the fitted oscillator parameters
are shown in the panels of Fig.~\ref{fig:eu1}.  While the two modes display a
strong polarization dependence, the modulation is not perfect.
The polarization modulation of the oscillators may be used to estimate the degree to
which the crystal is detwinned in the following way,
\begin{equation}
  \alpha_j = 1-\frac{\Omega_j^2(\perp)}{\Omega_j^2(\parallel)},
  \label{eq:alpha}
\end{equation}
where $\perp$ and $\parallel$ denote the polarizations perpendicular and parallel to the
dipole moment of the $j$th vibration.  In the case of a twinned crystal, there is no
polarization dependence, $\Omega_j(\perp) = \Omega_j(\parallel)$, and $\alpha_j=0$;
if the crystal is completely detwinned, then $\Omega_j(\perp)=0$ and $\alpha_j=1$.
Using the parameters for the $B_{2u}$ mode at 5~K yields $\alpha\simeq 0.72$; the
average for the two modes of $\alpha\simeq 0.7$ indicates that the crystal
is roughly 70\% twin free, which is comparable to the estimate of $\sim 80$\% based
on transport measurements \cite{He2017}.  We speculate that by gluing the GFRP and
sample arrangement to an optical mount, the strain imparted on the sample may be
slightly reduced, resulting in the slightly lower degree of detwinning observed in
this work.

%
%
\begin{figure}[tb]
\centerline{\includegraphics[width=3.3in]{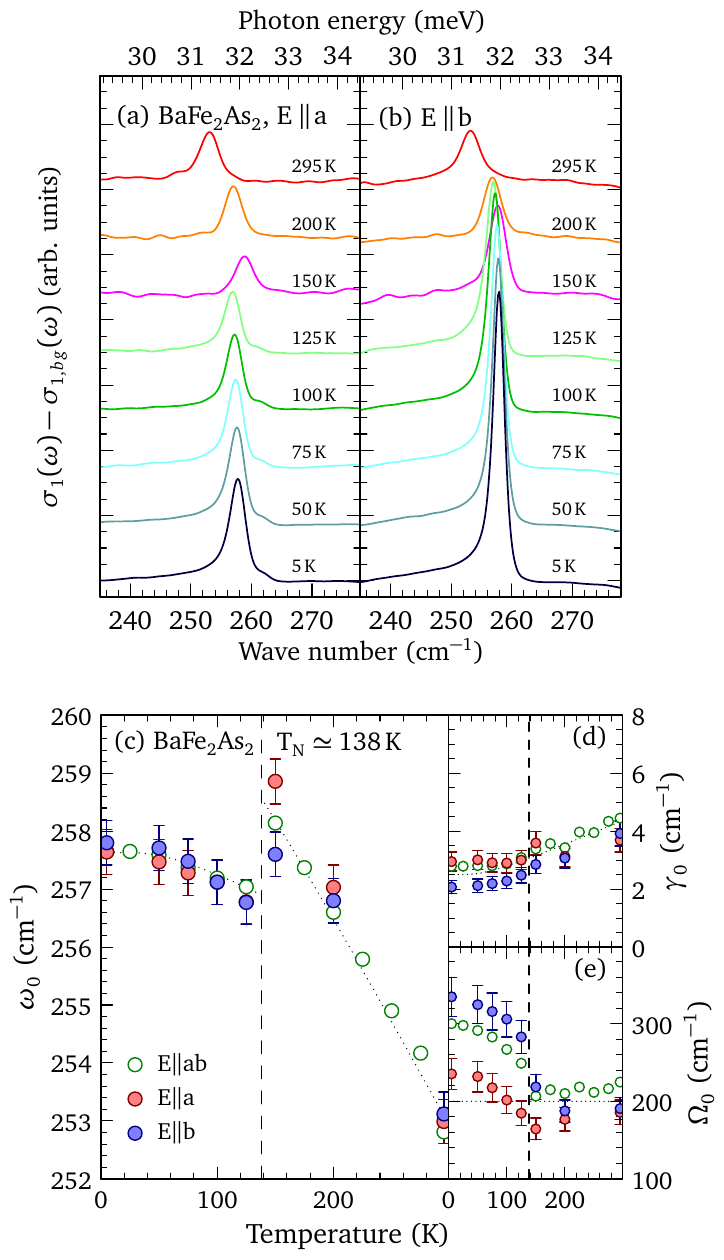}}%
\caption{The temperature dependence of the real part of the optical conductivity
in the region of the high-frequency $B_{3u}$ and $B_{2u}$ modes with the electronic background
removed for light polarized, (a) along the \emph{a} axis, and  (b) along the \emph{b}
axis; offsets have been added for clarity.  The temperature dependence of the (c) position,
(d) width, and (e) strength of the lattice mode in a twinned sample \cite{Homes2018},
compared to the polarized results determined in this work.}
\label{fig:eu2}
\end{figure}
%

%
%
%
\subsubsection{High-frequency mode}
The high-frequency $E_u$ mode is of considerable interest as one branch displays an anomalous
increase in oscillator strength below $T_{\rm N}$ \cite{Akrap2009}, with the other appears
to be largely silent \cite{Nakajima2011}.  The high-frequency mode has been fit using a
simple Lorentzian oscillator superimposed on a polynomial background; the resulting line shapes
are shown with the background removed for light polarized along the \emph{a} and \emph{b} axes
in Figs.~\ref{fig:eu2}(a) and \ref{fig:eu2}(b), respectively.  The line shape for this oscillator
at 5~K has an asymmetric line shape, suggesting electron-phonon coupling; while an asymmetric Fano
line shape has been fit to this vibration, the asymmetry parameter is very small \cite{Homes2018}
(see Fig.~S3 and the accompanying table).  The results for the position, width, and strength of the
oscillators along the \emph{a} and \emph{b} axis are shown in Figs.~\ref{fig:eu2}(c), \ref{fig:eu2}(d),
and \ref{fig:eu2}(e), respectively, where they are compared with results from a twinned sample
$({\rm E}\!\parallel\!{\rm ab})$ \cite{Homes2018}.
Above 200~K there is no polarization dependence in the oscillator parameters for this mode;
however, in a reversal of the behavior observed in the low-frequency $E_u$ mode, just above
$T_{\rm N}$ at 150~K the position of this mode for light polarized along what will become the
\emph{a} axis is slightly larger than it is along the \emph{b} axis, again indicating that
just above $T_{\rm N}$ the strain from the GFRP is a symmetry-breaking process and that the
lattice modes are likely coupling to the magnetism.
The frequency of this mode appears to decrease anomalously below $T_{\rm N}$; this is
understood as the splitting of the $E_u$ mode where the lower $B_{2u}$ branch is active
and the upper $B_{3u}$ branch is largely silent \cite{Homes2018}.  The fact that the positions
of the modes along the \emph{a} and \emph{b} axes are identical [Fig.~\ref{fig:eu2}(c)]
suggests that the activity of the mode along the \emph{a} axis is likely due to leakage
from the \emph{b} axis as a result of residual twins, and that the weak shoulder observed
just above this mode at $\simeq 261$~cm$^{-1}$ [Fig.~\ref{fig:eu2}(a) and supplementary
Fig.~S3(a)] is the $B_{3u}$ mode \cite{Schafgans2011}.
The oscillator strength of the $B_{2u}$ mode was observed to increase by a factor of two
in the twinned material \cite{Akrap2009}, but in the largely twin-free sample it has increased
threefold, in agreement with a previous study \cite{Nakajima2011}.  The origin of this anomalous
increase in the strength remains a topic of considerable debate.
Given the value of $\alpha\simeq 0.7$ determined from the modulation of the low-frequency
$E_u$ mode, and the value of $\Omega_0(\parallel)\simeq 340$~cm$^{-1}$ for the \emph{b}-axis
mode at 5~K, then the strength of the leakage should be $\Omega_0(\perp)\simeq 186$~cm$^{-1}$;
curiously, the observed value of $\Omega_0(\perp)\simeq 235$~cm$^{-1}$ is considerably larger
and suggests a lower value for the detwinning.  However, the rather peculiar nature of this mode
makes it a poor candidate for estimates of the degree of detwinning.  The more predictable
behavior of the low-frequency $B_{2u}$ and $B_{3u}$ modes suggests that the estimate of
$\simeq 70$\% detwinning is the more reliable one. \\

%
%
\section{Conclusions}
The optical properties of a large, detwinned single crystal of BaFe$_2$As$_2$ have
been determined above and below $T_{\rm N}\simeq 138$~K over a wide frequency range.
Above $T_{\rm N}$ the optical conductivity and the two infrared-active $E_u$ modes
are essentially isotropic; only the lattice modes display a weak polarization dependence
just above $T_{\rm N}$.
Below $T_{\rm N}$, the free-carrier response is strongly anisotropic; below $\simeq 30$~meV,
$\sigma_{1,a}/\sigma_{1,b}\sim 2$.  The narrow Drude component has only a weak polarization
dependence.  The anisotropy in the low-energy optical conductivity is driven by the difference
in the effective masses in the broad Drude component.  The interband contributions to the optical
conductivity above this energy appear to be isotropic.
The relatively large sample size allows the behavior of the lattice modes to be studied in detail.
The splitting of the low-energy $E_u$ mode is clearly observed, and the polarization modulation
of the resulting $B_{2u}$ and $B_{3u}$ modes are used to determine that the crystal is about
70\% detwinned.  The high-frequency mode remains enigmatic; the $B_{2u}$ component undergoes a
striking threefold increase in intensity, while the $B_{3u}$ mode is nearly silent.  This
relatively simple method for detwinning crystals may allow further detailed optical studies of
the nematic and (or) superconducting behavior in this family of materials.

%
%
%
\begin{acknowledgments}
We would like to acknowledge useful discussions with Ana Akrap and Mingquan He.
Work at Brookhaven National Laboratory was supported by the Office of Science, U.S.
Department of Energy under Contract No. DE-SC0012704.  Work at KIT was partially funded
by the Deutsche Forschungsgemeinschaft (DFG, German Research Foundation) - TRR 288 - 422213477
(project A2). %
\end{acknowledgments}

%
%
%
%

%

\hfill

%
%
%

\clearpage
\newpage

\newpage
\vspace*{-2.1cm}
\hspace*{-2.5cm}
{
  \centering
  \includegraphics[width=1.2\textwidth,page=1]{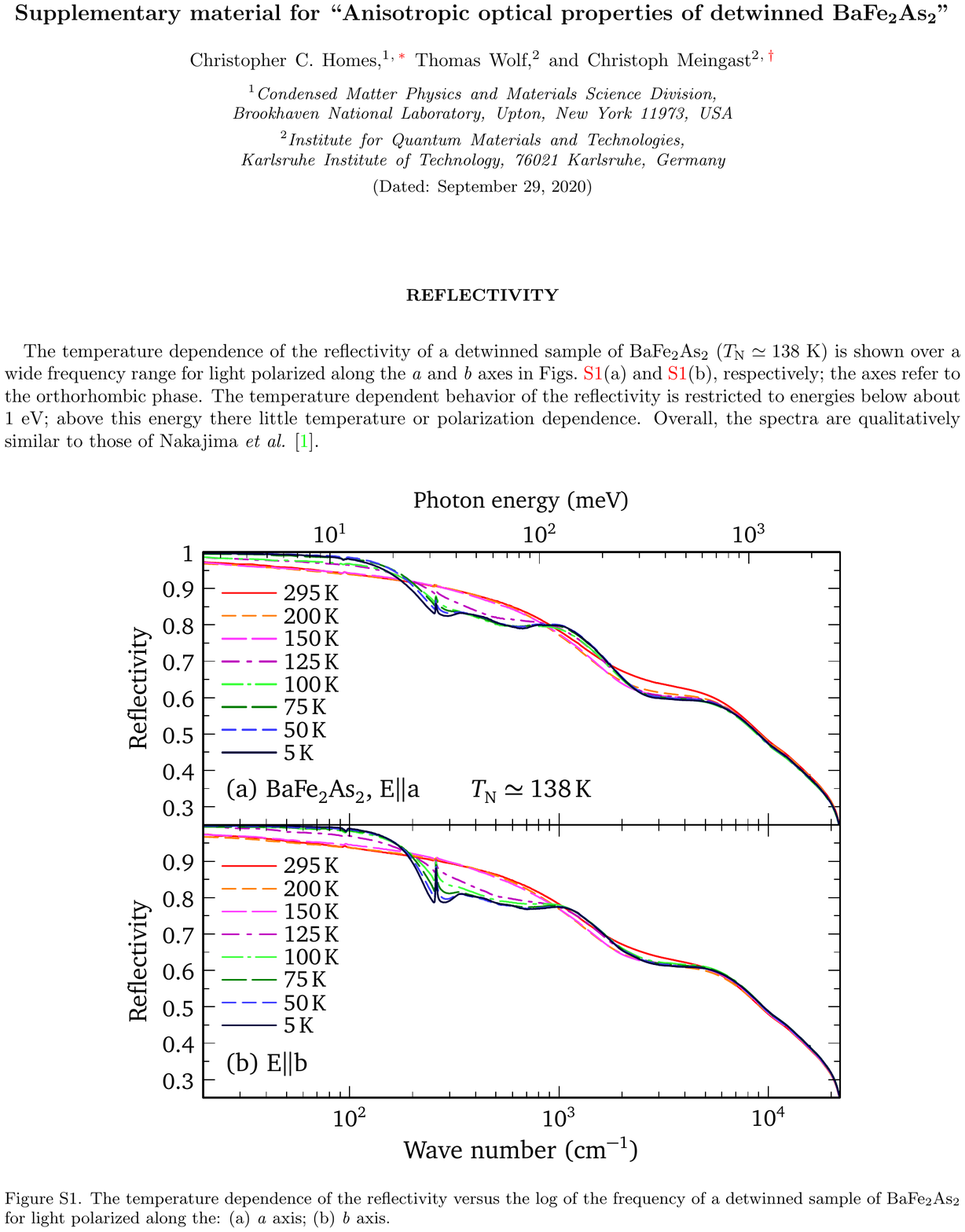} \\
  \ \\
}

\newpage
\vspace*{-2.1cm}
\hspace*{-2.5cm}
{
  \centering
  \includegraphics[width=1.2\textwidth,page=2]{supplemental.pdf} \\
  \ \\
}

\newpage
\vspace*{-2.1cm}
\hspace*{-2.5cm}
{
  \centering
  \includegraphics[width=1.2\textwidth,page=3]{supplemental.pdf} \\
  \ \\
}

\newpage
\vspace*{-2.1cm}
\hspace*{-2.5cm}
{
  \centering
  \includegraphics[width=1.2\textwidth,page=4]{supplemental.pdf} \\
  \ \\
}

\end{document}